\documentstyle[12pt]{article}
\begin{document}

\def\a{\alpha}
\def\b{\beta}
\def\ch{\chi}
\def\d{\delta}
\def\e{\epsilon}
\def\f{\phi}
\def\g{\gamma}
\def\h{\eta}
\def\i{\iota}
\def\j{\psi}
\def\k{\kappa}
\def\l{\lambda}
\def\m{\mu}
\def\n{\nu}
\def\o{\omega}
\def\p{\pi}
\def\q{\theta}
\def\r{\rho}
\def\s{\sigma}
\def\t{\tau}
\def\u{\upsilon}
\def\x{\xi}
\def\z{\zeta}
\def\D{\Delta}
\def\F{\Phi}
\def\G{\Gamma}
\def\J{\Psi}
\def\L{\Lambda}
\def\O{\Omega}
\def\P{\Pi}
\def\S{\Sigma}
\def\U{\Upsilon}
\def\X{\Xi}
\def\T{\Theta}

\def\xt{\tilde{x}}
\def\yt{\tilde{y}}
\def\zt{\tilde{z}}
\def\pp {\partial }
\hsize=6.in
\hoffset=-.3in
\textheight=8in

\thispagestyle{empty}
\baselineskip=7mm
 \begin{flushright} \ July \  1992 [Revised] \\ DAMTP R-92/24 \\ SNUTP 92-86
\end{flushright}
\begin{center}
\vglue .3in
{\bf  A Master Equation for  Multi-Dimensional Non-Linear Field Theories
}
\vglue .3in
 {\bf Q-Han Park}
\vglue .3in
{\it Department of Applied Mathematics and Theoretical Physics\\
University of Cambridge, Silver Street, \\
Cambridge, CB3 9EW, U.K. }
\vglue .2in
{ and }
\vglue .2in
{\it Department of  Physics,
Kyung Hee University\\
Seoul, 130-701, Korea }\footnote{
Address after Sept.1992; \hglue .1in E-mail address; qpark@nms.kyunghee.ac.kr }

\vglue .3in

{\bf ABSTRACT }
\vglue .3in
\end{center}
A master equation ( $n$ dimensional non--Abelian current conservation law
with mutually commuting current components ) is introduced for
multi-dimensional non-linear field theories. It is shown that the
master equation provides a systematic way to understand 2-d integrable
non-linear equations as well as 4-d self-dual equations and,
more importantly, their generalizations to higher dimensions.
\vglue .1in
\begin{center}
{To be published in Physics Letter B}
\end{center}
\vglue .1in

\newpage
\baselineskip=8mm

It is well known that 4-d self-dual equations (both the self-dual Einstein and
the self-dual Yang-Mills equations) are integrable via the twistor
method$^{[1][2]}$ and that after  dimensional reduction, they yield many known
integrable non-linear field equations in two spacetime dimensions.$^{[3][4]}$
Unfortunately, higher dimensional ($d \ge 3$) non-linear field equations,
describing non-trivial physical interactions do not arise in this way.
Moreover, it is not known whether there exist higher dimensional integrable
non-linear field equations similar to the two dimensional ones.

The purpose of this Letter is to introduce a set of simple first order
differential equations (the ``master equation") and show that it provides a
systematic way of obtaining the 4-d self-dual equations as well as generalizing
them to higher dimensions (the $2n$-d K\"{a}hler Ricci flat and the
Hermitian-Yang-Mills equations). When dimensionally reduced, these higher
dimensional ``self-dual" equations are shown to become various
multi-dimensional non-linear field equations, e.g. the 4-d sine-Gordon and the
4-d non-linear sigma model equations, and it is possible the master equation
share  the same integrability structure. Thus the master equation provides a
simple setting for the investigation of non-linear structures of higher
dimensional field equations, and may lead to a deeper understanding of the
subject. We discuss  the integrability of the master equation and consider the
possibility of multi-dimensional integrable field theory.

Consider the following $n$-dimensional equation (``master equation"):
\begin{equation}
(a) \ \ \nabla \cdot \mbox{\bf J}\equiv \pp_{i}J_{i} = 0  \ \ \  , \ \ \  \ \
(b) \ \ [ \ J_{i} \ , \ J_{j} \ ] = 0 \ \  ; \ \ i,j = 1,2, ... , n \ \ ( n \ge
2) \end{equation}
where $J_{i}$ are components of a vector current ${\bf J}$ valued in the Lie
algebra ${\bf g}$. The
bracket denotes a commutator and $\pp_{i} \equiv \pp / \pp x^{i} $ is the
partial differentiation with respect to local coordinates $x^{i}; i = 1,2, ...
, n$. We use the $n$-dimensional Euclidean metric to raise or lower indices.
Eq.(1) has a simple interpretation as a current
conservation law with all the current components mutually commuting. For $n =
2$,
the master equation becomes integrable and describes a 2-d
non-linear sigma model in the following way. Define the dual variable $A_{i}
\equiv \e_{ij}J_{j}$, with $\e_{12} = - \e_{21} = 1$,  to bring eq.(1) into the
form
 \begin{equation}
 a) \ \ \ \ \e_{ij}\pp_{i}A_{j} = 0 \  \ \ \ \ b)  \  \ \ \ \ [ \ A_{1} \ , \
A_{2} \ ] = 0 \ .
\end{equation}
 This may be solved partially for $A_{i} = g^{-1}\pp_{i}g $ with
$ g$ valued in the Lie group $G$. Then the remaining equation  becomes
\begin{equation}
\e_{ij}\pp_{i}(g^{-1}\pp_{j}g) = 0
\end{equation}
 which is precisely the equation of  the  2-d
sigma model with a pure Wess-Zumino term.$^{[5]} \ $ Eq.(2), on the other hand,
can be solved
completely  when we identify it as the integrability condition ( $ F_{12}^{\l }
\equiv [ L_{1}  , L_{2} ] = 0$ ) of the overdetermined
linear equations \begin{equation}
L_{i}\Phi \equiv (\pp_{i} + \l A_{i})\Phi = 0 \  \ \ \ \ ; \ \ \  i = 1,2
\end{equation}
for any value of  $\l \in CP^{1}$, and apply
the inverse scattering method$^{[6]}$ to the linear equations. In [5], we have
shown that the 2-d master equation ($n=2$) turns into
the 4-d self-dual Einstein and the self-dual Yang-Mills equations
when $J_{i}$  take values in the infinite
dimensional algebras $sdiffM^{2}$ and $sdiff_{ h }M^{2}$,\footnote{ Here
$sdiffM^{n}$ denotes the algebra of volume-preserving
diffeomorphisms of a manifold $M^{n}$ and $sdiff_{ h }M^{n}$ is its extension
by a Lie algebra ${\bf h}$.}  thereby relating 2-d integrable systems with 4-d
self-dual systems.

In this Letter,  we show that  this generalizes to
arbitrary $n$ subject  to certain subtleties given below. For $n \ge 2$, we
assume that $J_{i}$  take values in
$sdiffM^{n}$. Let $y^{\m } (\m = 1, \cdots , n)$ be coordinates on $M^{n}$. We
may write $J_{i}$ in a local coordinate basis\footnote{ We avoid possible
global problems and assume    that $M^{n}$ is simply connected and  vector
fields satisfy suitable regularity conditions.}$\{
 \pp_{\m } \equiv \pp / \pp y^{\m }  \ ; \ \m = 1,2,...,n \} $
\begin{equation}
J_{i} =
\e_{\m_{1} \cdots \m_{n}}\pp_{\m_{2}}V^{\m_{3} \cdots \m_{n}}_{i}
\pp_{\m_{1}}
\end{equation}
where $V^{\m_{1} \cdots \m_{n-2}}_{i} =
V^{[\m_{1}\cdots \m_{n-2}]}_{i} $ are
functions of the $2n$ variables $(x^{i}, y^{\m }  \ ; \ i,\m = 1,2,...,n \ )$.
Eq.(1a) can be solved by
\begin{equation}
V^{\m_{3} \cdots \m_{n}}_{i}  =\e_{ii_{2} \cdots i_{n}}\pp_{i_{2}}W^{\m_{3}
\cdots \m_{n}}_{i_{3}\cdots i_{n}}
\end{equation}
 with arbitrary functions $W^{\m_{3} \cdots \m_{n}}_{i_{3}\cdots i_{n}} =
W^{[\m_{3} \cdots \m_{n}]}_{[i_{3}\cdots i_{n}]}$. Note that eq.(5) is
invariant under the change
$W^{\m_{3} \cdots \m_{n}}_{i_{3}\cdots i_{n}} \rightarrow W^{\m_{3} \cdots
\m_{n}}_{i_{3}\cdots i_{n}} + \pp_{[i_{3}}F^{\m_{3} \cdots \m_{n}}_{i_{4}\cdots
i_{n}]}$
or $W^{\m_{3} \cdots \m_{n}}_{i_{3}\cdots i_{n}} \rightarrow W^{\m_{3} \cdots
\m_{n}}_{i_{3}\cdots i_{n}} + \pp^{[\m_{3}}G^{\m_{4} \cdots
\m_{n}]}_{i_{3}\cdots i_{n}}$
for arbitrary  $F$ and $G$. This shows that the $J_{i}$ are parameterized by
$(n-1)^{2}$ independent variables while the number of remaining equations in
eq.(1b) is ${1 \over 2}n^{2}(n-1)$.  Therefore the master equation is
overdetermined in general. A particular solution arises if we solve eq.(1)
partially
 in terms of a
scalar function $\O (x^{i},y^{\m}) $ such that
\begin{equation}
J_{i} =
\e_{\m_{1}\cdots \m_{n}}\e_{ii_{2} \cdots i_{n}}\O_{,\m_{2}i_{2}}\cdots
\O_{,\m_{n}i_{n}} \pp_{\m_{1}}
\end{equation}
where subscripts with a comma denote partial differentiation.
The remaining equations then become
\begin{equation}
[J_{i} \ , \ J_{j}] = \e_{\m_{1}\cdots \m_{n}}\e_{iji_{3}\cdots
i_{n}}\O_{,\m_{3}i_{3}}\cdots \O_{,\m_{n}i_{n}}D_{,\m_{2}}\pp_{\m_{1}} = 0
\end{equation}
where $D $ is the determinant of the $n \times n$ matrix whose  $(i, \m )$
entry is  $\O_{,i \m }$. Eq.(8) is satisfied if and only if $D_{,\m } =
\pp_{y^{\m }}D = 0$ for all $y^{\m }$. Therefore, after the integration, the
master equation reduces to
\begin{equation}
D = \e_{\m_{1}\cdots
 \m_{n}}\e_{i_{1}i_{2}\cdots i{n}}\O_{,\m_{1}i_{1}}\O_{,\m_{2}i_{2}}\cdots
\O_{,\m_{n}i_{n}}
  = \phi (x^{1}, \cdots x^{n})
\end{equation}
for an arbitrary function $\phi$. In the case $\phi \ne 0$, $\phi $ can be set
to one after a reparametrization of coordinates:
\begin{equation}
 (x^{i}, y^{\m }) \rightarrow (\tilde{x^{i}}, \tilde{y^{\m }});  \ \ \pp
\tilde{x^{1}} / \pp x^{1} = \phi , \ \ \tilde{x^{i}} = x^{i}; 2 \le i \le n, \
\ \tilde{y^{\m }} = y^{\m }; 1 \le \m \le n \ .
\end{equation}
 For $n=2$, eq.(7) is a general solution of eq.(1a) and so eq.(9) is in fact
equivalent to the master equation.  However,  for $n \ge 3$, it is not clear
whether eq.(9) is equivalent to the master equation or a special case of it
requiring further restrictions. This is an interesting open question which
will be pursued elsewhere. Here, we only wish to point out  that eq.(9) arises
from the master equation and that  eq.(9) has a simple   geometrical
interpretation which generalizes that given in [5] for the $n=2$ case. This is
as follows. Consider a K\"{a}hler metric given in terms of a potential $\O
$:\footnote{ \hsize=6in
We regard $ ( x^{i},y^{\m } )$ as $2n$ independent
complex variables so that we are really considering a complexified K\"{a}hler
structure. The usual   K\"{a}hler metric can be recovered by taking $y^{1} =
\bar{x^{1}}$ etc. We also use $w^{a}$ to denote the $2n$ coordinates $(x^{i},
y^{\m })$. }
 \begin{equation}
ds^{2} = g_{ab}dw^{a}dw^{b} =  2\O_{, i\m }dx^{i}dy^{\m }  \ \
. \end{equation} Then, eq.(9) with $\phi = 1$ means that the determinant of the
K\"{a}hler
metric is constant or, equivalently, that the Ricci tensor vanishes
identically, i.e.
$R_{ab} = 0$. Thus we have shown that  the $2n$-dimensional  K\"{a}hler Ricci
flat
equation arises from  eq.(1) with the infinite gauge symmetry
$sdiffM^{n}$. For $n = 2$, this is equivalent to the 4-d self-dual
Einstein equation while the linear equation (eq.(4)) determines
the corresponding twistor space.$^{[1][2]}$
 More generally, we may define a Lie algebra ${\bf h }$-extension of
$sdiffM^{n}$ by   the following exact sequence:
\begin{equation}
0 \rightarrow \mbox{\bf h} \rightarrow sdiff_{h}M^{n} \rightarrow sdiffM^{n}
\rightarrow 0  \ . \end{equation}
The algebra $sdiff_{h}M^{n}$ could be represented  in terms of ``covariantized"
 volume preserving vector
fields such that the $
J_{i} $ taking value in $sdiff_{h}M^{n}$ are given by
\begin{equation}
J_{i} =
\e_{\m_{1} \cdots \m_{n}}\pp_{\m_{2}}V^{\m_{3} \cdots \m_{n}}_{i}
\pp_{\m_{1}} + K_{i}
\end{equation}
where $K_{i} = K_{i}(x^{k},y^{\m})$ are  arbitrary functions valued in the Lie
algebra ${\bf h }$. It is easy to check that these covariantized  volume
preserving vector fields
close under commutation and indeed
 form an algebra. A partial solution of eq.(1) may once again  be obtained
in terms of a scalar function $\O $ satisfying eq.(9) and a holonomy element
$U$,
valued in the Lie group H associated with ${ \bf h}$, namely
\begin{equation}
J_{i} =
\e_{\m_{1}\cdots \m_{n}}\e_{ii_{2} \cdots i_{n}}\O_{,\m_{2}i_{2}}\cdots
\O_{,\m_{n}i_{n}} (\pp_{\m_{1}} + U^{-1}\pp_{\m_{1}}U) = { g^{i\m} \over
\sqrt{g}}
(\pp_{\m} +
U^{-1}\pp_{\m}U) \end{equation}
where $U$ satisfies
\begin{equation}
\e_{\m_{1}\cdots \m_{n}}\e_{ii_{2} \cdots i_{n}}(\O_{,\m_{2}i_{2}}\cdots
\O_{,\m_{n}i_{n}}  U^{-1}\pp_{\m_{1}}U)_{,i} =
 { g^{i\m} \over \sqrt{g}}\pp_{i}(U^{-1}\pp_{\m}U) = 0 \  . \end{equation}
Here $g^{ab}$ is the inverse of the metric given in eq.(11) and $g$ is the
determinant of the metric.
For $n=2$, this is precisely  Yang's expression for the 4-d self-dual
Yang-Mills equation on a
self-dual space.$^{[5]}$
 Thus eq.(15) generalizes the 4-d self-dual Yang-Mills equation
to the higher dimensional one known as  the $2n$-d
Hermitian-Yang-Mills equation:\footnote{
\hsize=6in
In general, Hermitian-Yang-Mills connections are given by eq.(16) and eq.(17)
but with $g^{i\m }F_{i\m } $ being a constant multiple of the identity. Here we
take the constant to be zero.}
 \begin{eqnarray}
F_{ij} &=& [{\pp \over \pp x^{i} }+A_{x^{i}} \ , \ {\pp \over \pp x^{j}} +
A_{x^{j}}] = 0
  \nonumber \\
F_{\m\n} &=& [{\pp \over \pp y^{\m} }+A_{y^{\m}} \ , \ {\pp \over \pp y^{\n}} +
A_{y^{\n}}] = 0  \\
g^{i\m}F_{i\m }  &=& g^{i\m} [{\pp \over \pp x^{i} }+A_{x^{i}} \  ,  \ {\pp
\over \pp y^{\m}} + A_{y^{\m}}] = 0 \
\end{eqnarray}
which can be shown to be equivalent to eq.(15) by solving eq.(16) for
$A_{x^{i}} = P^{-1}
\pp_{x^{i}}P, $  $  A_{y^{\m}} = Q^{-1} \pp_{y^{\m}} Q $ and rewriting
eq.(17) in terms of $U \equiv QP^{-1}$. Therefore, we have shown that the
master equation
systematically generalizes the 4-d self-dual
equations to higher dimensions.

 Our identification of the K\"{a}hler Ricci flat and the Hermitian-Yang-Mills
equations as the  proper higher-dimensional
generalization  of 4-d self-dual equations among other possibilities$^{[7][8]}$
stems from the following three observations, Firstly, they can be expressed  as
``generalized self-duality conditions" in $2n$
dimensions:$^{[10]}$\footnote{Eq.(19) for $n=3,4$ was first considered in [7]}
\begin{eqnarray} R_{abcd} &=& {}^{*}R_{abcd} \equiv {1 \over 2}\sqrt{g}
\e_{abefi_{1}\m_{1}\cdots i_{n-2}\m_{n-2}}g^{i_{1}\m_{1}}\cdots
g^{i_{n-2}\m_{n-2}}R_{cd}^{\ \ ef} \\ F_{ab}  &=& {}^{*}F_{ab} \ \ \ \equiv {1
\over 2}\sqrt{g} \e_{abcdi_{1}\m_{1}\cdots i_{n-2}\m_{n-2}}
g^{i_{1}\m_{1}}\cdots g^{i_{n-2}\m_{n-2}}F^{cd} \end{eqnarray} where
$\e_{ab\cdots cd} $ is the totally antisymmetric rank $2n$ tensor with
$\e_{x^{1}y^{1}\cdots x^{n}y^{n}} = 1$. Second, and more importantly,  the
K\"{a}hler Ricci flat and the Hermitian-Yang-Mills equations reduce to  many
interesting higher dimensional non-linear (interacting) field equations which
generalize 2-d integrable field equations. In the following, we give a few
examples. Consider eqs.(15)-(17) in the flat background given by $\O =
x^{1}y^{1} +
x^{2}y^{2} + \cdots + x^{n}y^{n} $. If we take $ x^{i} = \pm y^{i}$, eq.(15)
trivially reduces to the $n$-dimensional non-linear sigma model equation,
$\pp_{i}(U^{-1}\pp_{i}U) = 0$ with various signatures. For another example,
make
the  ansatz  $ A_{x^{n}} = \Psi  \ , \ A_{y^{n}} =
\Psi^{\dagger } $ with $\Psi $ in the adjoint representation and   $ y^{i} =
\bar{x}^{i} \  (i=1,2,...,n-1), $ where bar denotes  complex conjugation, and
where
all fields are independent of $x^{n}$ and $y^{n}$. Then
eqs.(16) and (17) become \begin{eqnarray} [ D_{x^{i}} \ , \ \Psi ] \equiv [
{\delta \over \delta  x^{i}} + A_{x^{i}} \ , \  \Psi ] = 0 \  & , &  \   [
D_{\bar{x}^{i}} \ , \ \Psi^{\dagger } ] \equiv [ {\delta \over \delta
\bar{x}^{i}}
 + A_{\bar{x}^{i}} \ , \ \Psi^{\dagger } ] = 0 \nonumber \\
\sum^{n-1}_{i = 1}F_{x^{i}\bar{x}^{i}} = - [ \Psi \ , \ \Psi^{\dagger } ] \ &
, & \
F_{x^{i}x^{j}} = 0 = F_{\bar{x}^{i}\bar{x}^{j}}
\end{eqnarray}
which imply the $(2n-2)$-d Yang-Mills-Higgs equations;
\begin{equation}
 a) \ \ \   [ D^{a} \ , \ [ D_{a} \ , \ \Psi ]] = 0 \ \ \ \ \ \ \
\ \ \  \ b) \ \  \  [ D^{a} \ , \ F_{ab} ] = [ D_{b} \ , \  [ \Psi \  , \
\Psi^{\dagger } ]] \ . \end{equation}
Now, restrict further by taking the following ansatz for $A_{x^{i}}$ and
$\Psi $:$^{[9]}$ \begin{equation}
A_{x^{i}} = \sum^{r}_{\a = 1}A_{x^{i}}^{\a }h^{\a } =
- \sum_{\b = 1}^{r}( K^{-1})_{\a\b }\pp_{x^{i}}\mbox{ \ ln  }u_{\b } h^{\a } \
; \
A_{\bar{x}^{i}} = - \sum^{r}_{\a = 1}(A_{x^{i}}^{\a })^{*}h^{\a }
\ ; \
\Psi = \sum^{r}_{\a = 1}u_{\a }e^{\a } \
\end{equation}
 where $h^{\a }, e^{\a }$ are Cartan-Weyl basis for the algebra satisfying
(without sum)
\begin{equation}
[e^{\a } \ , \ e^{-\a^{'}}] = \d_{\a\a^{'}}h^{\a } \ , \ \ \
[h^{\a } \ , \ e^{\b } ]    = K^{\b\a }e^{\b } \ , \ \ \
[h^{\a } \ , \ h^{\b } ] = 0 \ .
\end{equation}
Here $r$ is the rank of  the group and $K$ is the Cartan matrix. With $\phi_{\a
}
\equiv |u_{\a }|^{2}$, eq.(20) becomes the $(2n-2)$-d Toda equation;
\begin{equation}
\nabla^{2}ln\phi_{\a } = -\sum_{\b = 1}^{r}K_{\a\b }\phi_{\b }
\end{equation}
where $\nabla^{2}$ is the $(2n-2)$-d Laplacian. By using the affine
Cartan matrix for the Kac-Moody algebra, this can be generalized to the
affine Toda equation. In particular, for the $sl(2,R)$ Kac-Moody algebra, it
becomes the well-known $(2n-2)$-d sine(sinh)-Gordon equation.
We may also consider a reduction of the K\"{a}hler Ricci flat equation. Besides
the trivial
reduction to  the 4-d self-dual Einstein equation, it also reduces to less
trivial, non-self-dual equations. For example, assume a rotational Killing
symmetry
 s.t. $\O = \O (x^{i}, y^{i}, r \equiv
x^{n}y^{n}) ( i = 1,...,n-1)$ and make the change of coordinates: $(x^{i},
y^{i}, r) \rightarrow (p^{i} = x^{i}, q^{i} = y^{i}, w = r\pp_{r}\O ) $.
This allows us  to introduce a new K\"{a}hler potential $\S = \S
(p^{i},q^{i},w)$ satisfying  $\O_{,x^{i}} = \S_{,p^{i}}. $ The
reduced K\"{a}hler Ricci flat equation then becomes
\begin{equation}
det ( \S_{,p^{i}q^{j}} ) = (e^{-\S _{,w}})_{,w} \ .
\end{equation}
For $n=2$, eq.(25) was identified in [5] with the $sl(\infty )$-Toda equation.
However, this is not  in general true for $n \ge 3$. If we restrict further by
making the
following  separation of variables:
\begin{equation}
\Sigma (p^{i},q^{i},w)  = wK(p^{i},q^{i}) + f(w)
\end{equation}
we get the $(2n-2)$-d Einstein-K\"{a}hler equation:
\begin{equation}
det (g_{p^{i}q^{j}} ) = det (K_{,p^{i}q^{j}} ) = e^{\L K}
\end{equation}
with a cosmological constant $\L $. These examples show that most of the
four-dimensional extension of  two-dimensional relativistic integrable
equations can be embedded into the K\"{a}hler Ricci flat and the
Hermitian-Yang-Mills equations by almost the same reduction procedure as the
two dimensional case.
However, one should not that since for  $n \ge 3 $ eqs.(9) and (15) are not
necessarily equivalent to the master equation with the previously considered
infinite dimensional gauge groups, one may still need further specifications
for the precise ans\"{a}tze.
Third, we show that the Hermitian-Yang-Mills equations admit exact
multi-centered solutions in the particular case where the gauge group is chosen
to be $Sl(2,C)$. Consider the following ansatz for $A$ with a flat background
metric and  $y^{i} = \bar{x^{i}} ( i = 1, \cdots , n)$:
\begin{equation}
 A_{x^{i}} = \left( \matrix{ -{1\over 2}\pp_{i}ln \phi & - \pp_{i}ln \phi \cr
0 & {1\over 2}\pp_{i}ln \phi \cr}\right)
\ \ , \ \  A_{y^{\m }} = \left( \matrix{ {1\over 2}\pp_{\m }ln \phi &
                         0 \cr - \pp_{\m }ln \phi &
- {1\over 2}\pp_{\m }ln \phi \cr}\right) \  .
\end{equation}
Then, the  (complexified) Hermitian-Yang-Mills equations reduce to
\begin{equation}
{1 \over \phi }(\pp_{x^{1}}\pp_{\bar{x^{1}}}+ \cdots
+\pp_{x^{n}}\pp_{\bar{x^{n}}})\phi = {1 \over \phi }\nabla^{2} \phi = 0
\end{equation}
where $\nabla^{2}$ is the $2n$-d Laplacian.
This may be compared with  the 4-d $SU(2)$ self-dual  Yang-Mills  case
where  multi-centered  exact  solutions  with  $5k$  parameters   were
obtained by reducing the non-linear self-dual Yang-Mills  equation  to
the linear  4-d  Laplace  equation.$^{[11]}$  However,  it  should  be
emphasized that our ansatz is different from those in  [11]  and  does
not apply to the $SU(2)$  case,  and  moreover,  makes  the  Euclidean
Yang-Mills  action  vanish  identically.  The  explicit  form  of  the
solutions and further details will be given in [10]. Unfortunately, we
have not succeeded in obtaining exact multi-centered solutions for the
$SU(2)$ case and other Hermitian cases. Nevertheless, there  exists  a
theorem proven by Uhlenbeck and  Yau$^{[12]}$  which  shows  that  the
moduli space of $2n$-d Hermitian-Yang-Mills connections  is  identical
to that of stable holomorphic vector bundles over a compact K\"{a}hler
manifold.  This theorem generalizes the theorem of Donaldson  for  the
$n=2$ case$^{[13]}$ which simplifies the ADHM construction$^{[14]}$ of
Yang-Mills instantons. Even though   the Uhlenbeck and  Yau's  theorem
is not as strong as  the  ADHM  construction  to  allow  the  explicit
construction of solutions (it is only an existence proof), it provides
a  certain  algebraic  setting  for  the  problem  and  supports   our
identification of the  Hermitian-Yang-Mills  equation  as  the  proper
higher dimensional generalization  of  the  4-d  self-dual  Yang-Mills
equation.

Having shown that the master equation  implies  various multi-dimensional
non-linear field equations which become integrable in two
spacetime dimensions, we are now led to the question as to whether these
non-linear equations are, in general, integrable. In the $n
 = 2$ case, both the master equation and  4-d self-dual equations (as well as
their 2-d reductions) possess the same integrability
structure, i.e. they all appear to be the integrability conditions of  the
associated
linear systems (eq.(3)), while general solutions are obtained by solving the
Riemann-Hilbert problem corresponding to linear systems.$^{[1][2][6]}$ This
leads us to conjecture that the integrability structure of the K\"{a}hler Ricci
flat and the Hermitian-Yang-Mills equations as well as their various reductions
is essentially the same as that of the master equation. At present, it is not
known whether the master equation for $ n \ge 3$ is integrable. If so, it would
possibly require a new notion of integrability (e.g. flatness associated with
the motion of extended objects) and an introduction of higher order non-linear
cohomology which would eventually lead to multi- dimensional integrable field
theory. This issue will be considered elsewhere.$^{[10]}$ \vglue .3in

{\bf ACKNOWLEDGEMENTS }

\vglue .2in

I would like to thank M.F.Atiyah, N.J.Hitchin for explaining their unpublished
work and initiating my interest in 6-d gauge theories. I am also indebted to
G.Gibbons, P.Goddard, P.K.Townsend, P.S.Howe, M.Blencowe and
S.Majid for discussions and T.Samols for his reading of the manuscript. I am
grateful to CTP at Seoul National University, RIMS at Kyoto
University for their help during my visit,  and above all to S.Hawking and
K.Soh
for their encouragement.

\def\Item{\par\hang\textindent}

{\bf REFERENCES }
\Item {[1]} R.Penrose, Gen.Rel.Grav.{\bf 7} (1976) 31
\Item {[2]} R.S.Ward,Phys.Lett.{\bf
61A } (1977) 81;  M.F.Atiyah and R.S.Ward, Commun.Math.Phys.{\bf 55}, (1977)
117
\Item {[3]} R.S.Ward,Phil.Trans.Roy.Soc.Lond. {\bf A315} (1985) 451;
N.J.Hitchin, Proc.Lond.Math. Soc. {\bf 55}  (1987) 59
\Item {[4]} I.M.Krichever and S.P.Novikov, Funkts.Analiz.12(4) (1978) 41;
L.J.Mason and G.A.J.Sparling, Phys.Lett.{\bf 137A} (1989)
29; I.Bakas and D.A.Depireux, Mod.Phys.Lett. {\bf A6} 399(1991)
\Item {[5]} Q-H.Park, Phys.Lett.{\bf 238B},287 (1990); Phys.Lett.{\bf257B},
105(1991);  Int.J.Mod.Phys.A, Vol. 7,No. 7 (1992) 1415
\Item {[6]} C.S.Gardner,I.M.Greene,M.D.Kruskal and R.M.Miura, Phys.Rev.Lett.
{\bf 19} (1967) 1095; V.E.Zakharov and A.B.Shabat, Funct.Anal. and Appl. {\bf
13} (1979) 13; V.E.Zakharov and A.V.Mihailov, Sov.Phys.JETP {\bf 47} (1978)
1017
\Item {[7]} E.Corrigan,C.Devchand,D.B.Fairlie and J.Nuyts, Nucl.Phys.{\bf B214
} (1983) 452;  R.S.ward, Nucl.Phys. {\bf B236} (1984) 381
\Item {[8]} D.H.Tchrakian, J.Math.Phys. {\bf 21} (1980) 166
\Item {[9]} A.N.Leznov and M.V.Saveliev, Commun.Math.Phys. {\bf 74 }(1980) 111
\Item {[10]} Q-H. Park, ``On Multi-Dimensional Integrable Field Theory", in
preparation
\Item {[11]} G.t'Hooft (unpublished); R.Jackiw, C.Nohl and C.Rebbi,
Phys.Rev.{\bf D15}, (1977) 1642
\Item {[12]} K.K.Uhlenbeck and S.-T. Yau, Commun.Pure.Appl.Math.{\bf 39}(1986)
257; correction, ibid {\bf 42},703
\Item {[13]} S.K.Donaldson,  Proc.Lond.Math.Soc.{\bf 3}(1985) 1
\Item {[14]} M.F.Atiyah, V.G.Drinfeld, N.J.Hitchin and Y.I.Manin,
Phys.Lett.{\bf A65} (1978) 185

\end{document}